# PRACTICAL AND ALGORITHMICAL MANIFESTATIONS OF QUANTUM CHAOS*

Baowen Li and Marko Robnik
*Center for Applied Mathematics and Theoretical Physics,*
*University of Maribor, Krekova 2, SLO-62000 Maribor, Slovenia*
E-mail: Baowen.Li@uni-mb.si,  Robnik@uni-mb.si

## ABSTRACT

Quantum chaos manifests itself also in algorithmical complexity of methods, including the numerical ones, in solving the Schrödinger equation. In this contribution we address the problem of calculating the eigenenergies and the eigenstates by various numerical methods applied to 2-dim generic billiard systems. In particular we analyze the dependence of the accuracy (errors) on the density of discretization of the given numerical method. We do this for several different billiard shapes, especially for the model billiard system inside the region defined by quadratic (complex) conformal map $w = z + \lambda z^2$ of the unit disk $|z| \leq 1$ as introduced by Robnik (1983, 1984). As the shape parameter $\lambda$ varies within the interval $[0, 1/2]$ the system goes from integrable circle billiard ($\lambda = 0$), through a KAM-like regime ($0 < \lambda < 1/4$), and through an almost ergodic regime of hard chaos becoming rigorously ergodic, mixing and K-system at latest for $\lambda = 1/2$, as has been recently rigorously proved by Markarian (1993). We study the numerical error of the boundary integral method and the plane wave decomposition method (compared to the "exact" numerical results by the conformal mapping diagonalization technique obtained by Prosen and Robnik (1993,1994)) as a function of the discretization parameter $b$ which by definition is the number of discretization nodes on the boundary per one de Broglie wavelength (arclength) interval. For boundary integral method, we discover that at each $\lambda$ the error scales as a power law $< |\Delta E| > = Ab^{-\alpha}$, where $\alpha$ is a strong function of $\lambda$: In the KAM-like regime $0 \leq \lambda \leq 1/4$ it is large and close to 3.5, but close to $\lambda = 1/4$ it changes almost discontinuously becoming hardly any larger than zero. This is because the billiard becomes nonconvex beginning at $\lambda = 1/4$. For the plane wave decomposition method, we found at a fixed $b$ that the average absolute value of the error $< |\Delta E| >$ correlates strongly with the classical chaos, where the error $< |\Delta E| >$ does increase sharply with increasing classical chaos. This property is valid also for Bunimovich stadium and the Sinai billiard.

## 1. Introduction

In the studies of quantum chaos (Gutzwiller 1990, Giannoni *et al* 1991, Chirikov and Casati 1995) good numerical methods are not only indispensable but quite essential. We need them in order to illustrate and verify the theoretical developments.

---

Also, new numerical experiments provide valuable material as evidence and inspiration for new theories. Quantum billiards are certainly very useful model systems since dynamically they are generic and - depending on the design - they can cover all regimes of classical motion between integrability and full chaos (ergodicity). Their classical dynamics can be easily followed for very long time periods because no integration is necessary but only searching for zeros of certain functions at collision points. Quantally they have the advantage of having a compact configuration space admitting application of various numerical methods such as the Boundary Integral Method (BIM) (Banerjee 1994, Berry and Wilkinson 1984), the Plane Wave Decomposition Method (PWDM) employed by Heller (1984, 1991), conformal mapping diagonalization technique introduced by Robnik (1984), recently by Prosen and Robnik (1993). Among the general methods BIM is probably most widely used, even in quite practical engineering problems.

The main task of the present paper is to present our recent results on BIM and PWDM (Li and Robnik 1995 b,c). Especially, we want to discuss problems in BIM connected with the nonconvex geometries. It can be shown and qualitatively explained, using semiclassical arguments, that the usual and naive implementation of this method meets severe difficulties or even completely fails in nonconvex geometries. Otherwise the difficulties (bad convergence with the increasing density of discretization $b$) do not correlate with classical chaos.

This aspect is quite different in PWDM, which always works brilliantly in integrable systems (very fast convergence with $b$), but at fixed $b$ the accuracy degrades dramatically with increasing degree of classical chaos. This phenomenon can be qualitatively understood by the observation that in the semiclassical limit the wavefunction can be locally represented by a superposition of a finite number of plane waves, whereas in the classically chaotic (ergodic) system we need locally infinitely many plane waves. Therefore in a mixed system of KAM type we expect substantial degradation of the accuracy (at fixed $b$) with increasing size of the chaotic component. This is exactly what we find and report in section 3.

We believe - and this is the subject of our recent paper (Li and Robnik 1995d) - that the above manifestation of quantum chaos in applications of numerical methods are deeply connected to the notion of *sensitivity of the eigenfunctions* of classicaly chaotic quantal Hamiltonian systems with respect to the boundary conditions, and/or boundary data and/or the system parameter. It is expected that such sensitivity would indeed correlate with classical chaos whereas in the classicaly integrable systems it would be lacking. It was the important pioneering idea of Percival (1973), based on the semiclassical thinking, that proposed to classify the eigenstates and the energy levels in regular and irregular depending on whether they are associated with (supported by) classical regular regions (invariant tori) or by classical chaotic regions in the classical phase space. This picture is in fact basis of the Berry-Robnik (1984) approach to describe the statistical properties of energy spectra in the transition region between integrability and full chaos (ergodicity), which has been recently fully confirmed (Prosen and Robnik 1994 a,b). Moreover, we have recently performed the first *dynamical* separation of regular and irregular levels and eigenstates (Li and Robnik 1995a, 1994), thereby explicitly verifying the ingredients of the Berry-Robnik picture. We will discuss the quantitative aspects of sensitivity in the sense of the level curvature distribution in the final section.

## 2. Boundary Integral Method

We are searching for the solution $\psi(\mathbf{r})$ with eigenenergy $E = k^2$ obeying the Helmholtz equation

$$\nabla_{\mathbf{r}}^2 \psi(\mathbf{r}) + k^2 \psi(\mathbf{r}) = 0, \tag{1}$$

with the Dirichlet boundary condition $\psi(\mathbf{r}) = 0$ on the boundary $\mathbf{r} \in \partial \mathcal{B}$. It is possible to transform this Schrödinger equation to an integral equation which can be discretized and numerically solved. The main parameter here is the density of the discretization $b$ defined as the number of numerical nodes per one de Broglie wavelength along the boundary (assuming equidistant locations). Thus $b = 2\pi N/(k\mathcal{L})$, where $N$ is the total number of numerical nodes on the boundary whose perimeter is $\mathcal{L}$.

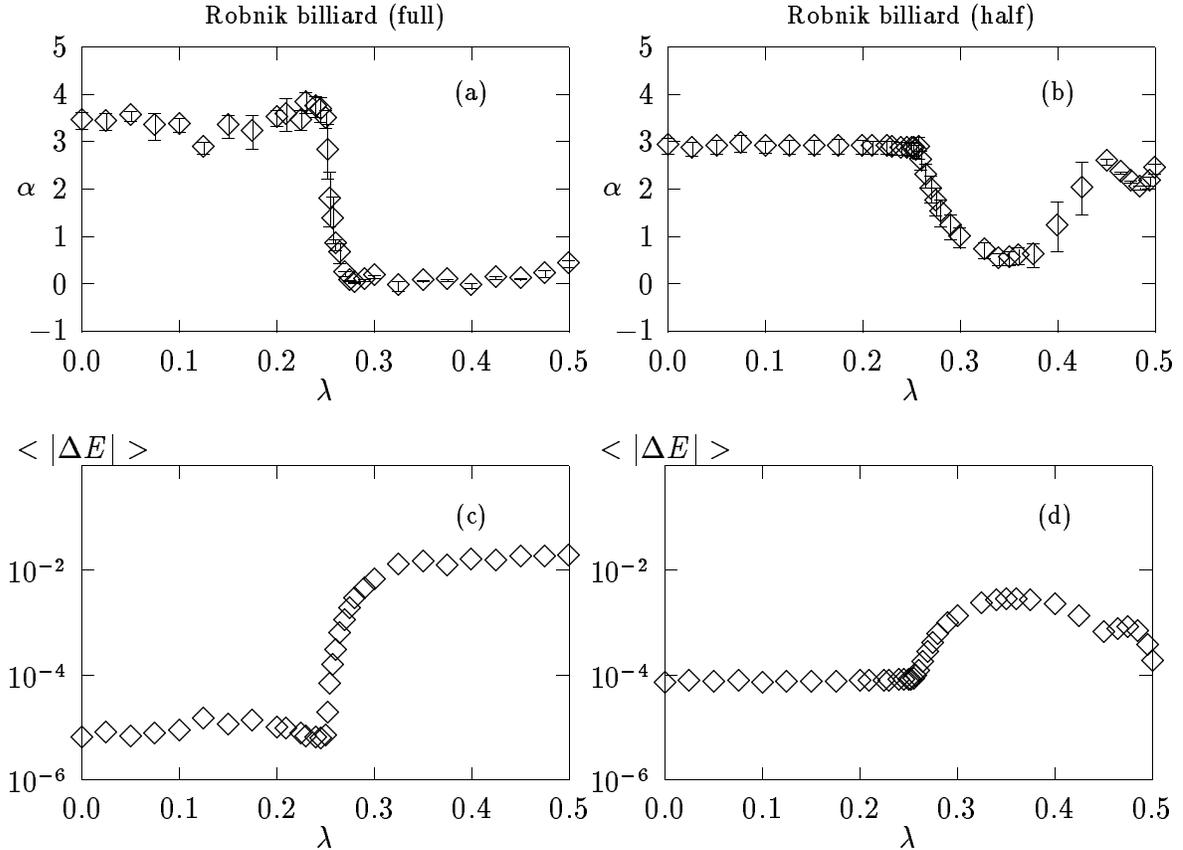

Fig. 1. $\alpha$ versus $\lambda$ plot for Robnik billiard in (a) and for half Robnik billiard in (b). In (c) and (d) we plot $\lg(<|\Delta E|>)$ with fixed $b = 12$ versus $\lambda$ for the two billiards, respectively.

We have investigated the dependence of the error $\Delta E$ as a function of $b$ for a variety of billiards. Of course $\Delta E$ fluctuates wildly from state to state and therefore some averaging is necessary. We have always taken the average over the lowest 100 eigenstates and confirmed carefully that this mean value is sufficiently stable (stationary) and thus well defined. Also, from now on it will be assumed that the energy and $\Delta E$ are measured in the units of the mean level spacing. By $<|\Delta E|>$ we shall denote the average of $|\Delta E|$ (in units of the mean level spacing) over the lowest 100 states. Our main result is that there is always a power law $<|\Delta E|>= Ab^{-\alpha}$ but the exponent $\alpha$ is nonuniversal and typically becomes almost zero if there are nonconvex segments on the boundary, implying bad convergence then. The results are summarized in table 1 and in figure 1. The main conclusion is that $\alpha$ does not correlate with the degree of classical chaos but only with existence or nonexistence of nonconvex segments on the boundary.

Table 1. The power law exponent $\alpha$ and the average absolute value of the error $<|\Delta E|>$ with $b = 12$ for different billiards. For details of the KAM-type see also figures 1(a,b).

| Type | Quantum billiard | $\alpha$ | $<|\Delta E|>_{b=12}$ |
|---|---|---|---|
| Integrable | Circle (half) | $2.94 \pm 0.17$ | $6.74\text{E}-5$ |
| | Circle (full) | $3.44 \pm 0.18$ | $5.97\text{E}-6$ |
| | Rectangle-triangle | $3.28 \pm 0.29$ | $4.08\text{E}-5$ |
| KAM | Robnik (full) ($0 < \lambda < 1/4$) | $\approx 3.4$ | $\approx 5.0\text{E}-6$ |
| | Robnik (half) ($0 < \lambda < 1/4$) | $\approx 2.9$ | $\approx 7.0\text{E}-5$ |
| Ergodic | Stadium (1/4) | $3.00 \pm 0.16$ | $1.18\text{E}-4$ |
| | Cardioid (half) | $2.42 \pm 0.11$ | $1.76\text{E}-4$ |
| | Cardioid (full) | $0.42 \pm 0.08$ | $1.85\text{E}-2$ |
| | Sinai (1/8) | $-0.34 \pm 0.11$ | $3.63\text{E}-1$ |
| | Robnik (full) ($0.3 < \lambda < 1/2$) | see figure 1a | see figure 1c |
| | Robnik (half) ($0.3 < \lambda < 1/2$) | see figure 1b | see figure 1d |

### 3. Plane Wave Decomposition Method of Heller

In this section we shall analyze the behaviour of the PWDM of Heller (1984,1991) to solve the Helmholtz equation (1) using the *Ansatz* of the following superposition of plane waves

$$\psi(\mathbf{r}) = \sum_{j=1}^{N} a_j \cos(k_{xj} x + k_{yj} y + \phi_j), \qquad (2)$$

where $k_{xj} = k \cos\theta_j$, $k_{yj} = k \sin\theta_j$, $k^2 = E$, and we use the notation $\mathbf{r} = (x, y)$. $N$ is the number of plane waves and $\phi_j$ are *random phases*, drawn from the interval $[0, 2\pi)$, assuming uniform distribution, and $\theta_j = 2j\pi/N$ determining the direction angles of the wavevectors chosen equidistantly. The *Ansatz* (2) solves the Schrödinger equation (1) in the interior of the billiard region $\mathcal{B}$, so that we have only to satisfy the Dirichlet boundary condition. Taking the random phases, as we discovered, is equivalent to spreading the origins of plane waves all over the billiard region $\mathcal{B}$, and at the same time this results in reducing the CPU-time by almost a factor of ten. All technical details are given in (Li and Robnik 1995c) and due to the lack of space we here just discuss the results.

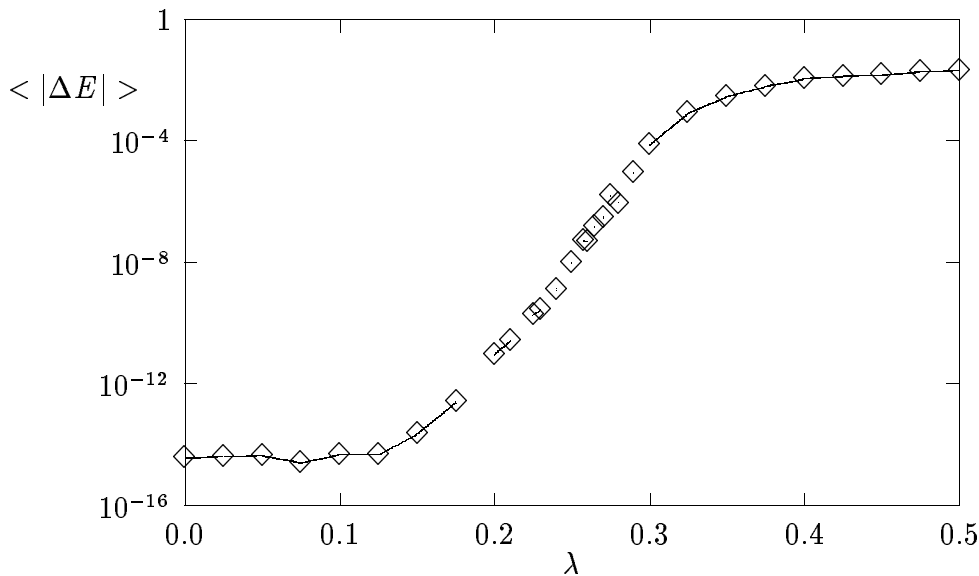

Fig. 2. The ensemble averaged (over 100 lowest odd eigenstates) absolute error (in units of mean level spacing) versus the billiard shape parameter $\lambda$ for Robnik billiard with $b = 12$. The numerical points are denoted by the diamonds, which are joined by straight lines just to guide the eye.

Unlike in BIM in using this method we find strong correlation between the accuracy (at fixed $b$), or the error $< |\Delta E| >$, with degree of classical chaos. This is clearly demonstrated in figure 2 for the Robnik billiard (conformal quadratic map $w = z + \lambda z^2$ from the unit disk $|z| \leq 1$ onto the $w$ plane) where at fixed $b = 12$ we observe continuous transition of the error from almost double precision (14 valid digits) in the integrable case of circle billiard ($\lambda = 0$), to the fully ergodic case of $\lambda = 1/2$, where the error is just a few percent. To demonstrate our qualitative assertion we have also looked at the Sinai billiard and the stadium billiard of Bunimovich. In both cases we found instantaneous drop of accuracy with even a very small deformation of the integrable billiard which in both cases becomes ergodic at any size of the deformation. These results certainly agree with our predictions and expectations expounded in the introduction.

## 4. Discussions and conclusions

As outlined in the introduction we have demonstrated that the existence of clas-sical chaos implies loss of accuracy of quantal numerical methods to solve a quantum billiard for some general methods like PWDM of Heller, but not in BIM. This behaviour has been qualitatively explained in the introduction. Here we just comment on relation between this observation and the notion of sensitivity with respect to the system parameter. Following the pioneering idea of Percival (1973) we are now able to understand this property in quantitative terms, namely in terms of energy level curvature distribution (of the unfolded spectrum). This distribution is delta function at zero curvature for integrable systems and widens with increasing degree of classical chaos. Although globally the distribution has no universality it has nevertheless universal tails. More details are given in our recent paper (Li and Robnik 1995d). When starting this project we were thinking quite generally about the notion of sensitivity of eigenstates introduced above and tried to implement some of these ideas also numerically in the sense of looking at the norm of the second derivatives of the wavefunctions, but we had to abandon this analysis simply because it is numerically too massive even for the lowest states. Thus as a partial substitute we looked in detail at the energy level flow and the curvature distribution. When setting up a proper theory we expect to find a deep connection between these phenomena, both being a manifestation of the sensitivity of eigenstates in presence of classical chaos.